\documentclass[preprint,prd,noshowpacs,nofootinbib]{revtex4}

\usepackage{color} 
\usepackage{amssymb}
\usepackage{amsmath}
\usepackage{graphicx} 
\usepackage{float}
\usepackage{braket}     
\usepackage{pdfpages}
\usepackage{epstopdf}
\usepackage[utf8]{inputenc}

\begin{document}

\title{Reconciling a quantum gravity minimal length with lack of photon dispersion}

\author{Michael Bishop}
\email{mibishop@mail.fresnostate.edu}
\affiliation{Mathematics Department, California State University Fresno, Fresno, CA 93740}

\author{Joey Contreras}
\email{mkfetch@mail.fresnostate.edu}
\affiliation{Physics Department, California State University Fresno, Fresno, CA 93740}

\author{Jaeyeong Lee}
\email{yeong0219@mail.fresnostate.edu}
\affiliation{Physics Department, California State University Fresno, Fresno, CA 93740}

\author{Douglas Singleton}
\email{dougs@mail.fresnostate.edu}
\affiliation{Physics Department, California State University Fresno, Fresno, CA 93740}

\date{\today}

\begin{abstract}
        Generic arguments lead to the idea that quantum gravity has a minimal length scale. A possible observational signal of such a minimal length scale is that photons should exhibit dispersion. In 2009, the observation of a short gamma ray burst seemed to push the minimal length scale to distances smaller than the Planck length. This poses a challenge for such minimal distance models. Here we propose a modification of the position and momentum operators, ${\hat x}$ and ${\hat p}$, which lead to a minimal length scale, but preserve the photon energy-momentum relationship $E = p c$. In this way there is no dispersion of photons with different energies. Additionally, this can be accomplished without modifying the commutation relationship $[{\hat x}, {\hat p}] = i \hbar$. 
\end{abstract}

\maketitle

\section{Gamma ray burst constraints on minimal length}

Theories of quantum gravity often incorporate a minimum distance at very high energy/momentum scales \cite{garay}. One argument for such a minimum length \cite{scardigli} is that at low energy/momentum scales the relationship between momentum uncertainty is given by the usual Heisenberg uncertainty principle, namely $\Delta x \approx \frac{\hbar}{\Delta p}$. To probe smaller $\Delta x$ one needs to collide particles at higher energy/momentum. Eventually, one forms a micro-black hole and thereafter increasing the energy/momentum of the collision only increases (linearly) the Schwarzschild radius of the micro-black hole. In mathematical terms $\Delta x \approx 2 G M \to 2 G \Delta p$, so that as $\Delta p$ increases so does $\Delta x$. In the above we have taken $c=1$. Combining both regimes gives $\Delta x \approx \frac{\hbar}{\Delta p} + 2 G \Delta p$ which yields a non-zero minimum for $\Delta x$.

There are different laboratory methods (Lamb shift, scanning tunneling microscope, electroweak precision measurements, Landau levels, evolution of nano-scale oscillators) to test for such a minimal length \cite{vagenas,vagenas2,bawaj}. There are also more recent proposals to use gravitational waves \cite{bosso} or the spreading of the wavefunction of large molecules \cite{sujoy,sujoy2,sujoy3} as ways to test for a minimum length. Up to date limits of the various bounds on a quantum gravity minimal length coming from different laboratory experiments can be found in \cite{scar-2015,lamb}.  Some of the most stringent constraints on a minimal length comes from astrophysical observations. One particular proposal of this type \cite{AC-nature,AC-PLB,AC-ijmpd} is to test for a minimal length scale using observations of gamma ray bursts (GRB). Gamma ray bursts are emissions of extremely high energy photons that generally are detected after traveling large, cosmological distances. Gamma ray bursts fall into two categories: (i) short gamma ray bursts which are thought to come from neutron star mergers or neutron star-black hole mergers and (ii) long gamma ray bursts which are thought to come from supernova. It is the short gamma ray bursts which are most useful in potentially observing the effects of a minimal length scale. As photons travel from the GRB to Earth, they should generically exhibit dispersion due to a minimal length {\it i.e.} photons of different energies will have slightly different velocities.  
 
In references \cite{AC-nature,AC-PLB,AC-ijmpd} a generic quantum gravity modified energy-momentum relation for photons was proposed of the form
\begin{equation}
    \label{AC E'}
    p^2 c^2 = E^2 [1 + f\left(E/E_{QG}\right)] ~,
\end{equation}
where $f(E/E_{QG})$ is some arbitrary function associated with one's theory of quantum gravity and with $E_{QG}$ being the energy scale of quantum gravity. Often this is set to be the Planck scale ($E_{QG} = E_{Pl} = \sqrt{\hbar c^5/G} \approx 10^{19}$ GeV).  When $E \ll E_{QG}$, the Taylor expansion of \eqref{AC E'} leads to
\begin{equation}
\label{AC E'1}
    p^2 c^2 = E^2 [1 + \xi (E/E_{QG}) + {\cal O} (E/E_{QG})^2 ].
\end{equation}
The sign of $\xi$ is model dependent. From Hamilton's equation, the photon's velocity is given as $\frac{\partial E}{\partial p} = v$. Using \eqref{AC E'1} and series expanding gives an energy dependent photon velocity 
\begin{equation}
    \label{v}
    v = \frac{\partial E}{\partial p} \approx c \left(1 - \xi \frac{E}{E_{QG}} \right).
\end{equation}
This energy dependence of photon velocity leads to a difference in the arrival time, $\delta t$, of the photons with energy difference $\delta E$. To first order in $E$ this time delay is
\begin{equation}
    \label{delta t}
    \delta t = \xi \frac{L}{c}\frac{\delta E}{E_{QG}} ~,
\end{equation}
where $L$ is the distance the photons traveled.
This implies that the farther photons travel and the larger the energy differences, the bigger the time delay. One of the goals of this work is to show that it is possible to have a minimal length scale while avoiding dispersion of the form in \eqref{v}.   

In 2009 \cite{abdo}, the Fermi Gamma-Ray space telescope detected a powerful, short GRB (GRB090510). The observations of GRB090510 allowed one to constrain the quantum gravity length scale from 1.2 to 100 times smaller the Planck length depending on whether one made conservative or liberal assumptions ({\it i.e.}  $L_{QG} \le \frac{L_{Pl}}{1.2}$ to $L_{QG} \le \frac{L_{Pl}}{100}$ or in terms of the Planck mass one has the inverse relationship $M_{QG} \ge 1.2 M_{Pl}$ to $M_{QG} \ge 100 M_{Pl}$). One of the assumptions that went into setting these bounds is that the lowest order correction to the dispersion relationship from \eqref{AC E'} was linear in $E$ as in \eqref{AC E'1}. If the lowest order correction is of order $E^2$ or higher then the constraints are much weaker. This result appeared to contradict the expectation that one should find evidence for a minimal length scale well before reaching the Planck scale. If taken at face value these results present a challenge to the ideas of minimal distance scales emerging from quantum gravity.

Modified dispersion relationships of the form in \eqref{AC E'} can be obtained from modifications of special relativity called deformed special relativity or double special relativity (DSR) \cite{AC-PLB,AC-ijmpd,dSR,dSR1,glikman,smolin}. DSR contains the idea of an observer independent minimal length. In this work we will approach a minimal length through a modification of the quantum mechanical position and momentum operators and their commutator. This approach is known as the generalized uncertainty principle (GUP) \cite{KMM}. Actually, the approach in this paper is slightly different from GUP since we do modify the position and momentum operators, but in such a way that the form of the commutator remains the same. 

Below we will present a minimal length scale model which does not lead to an energy dependent speed of light and thus avoids the constraints coming from the GRB observations. We propose {\it both} a modified position operator and modified momentum operator which lead to a minimal length but do so without leading to dispersion of photons of different energies. Many approaches to GUP modify only the position operator, as in \cite{KMM}, or only the momentum operator, as in \cite{vagenas2}. Our motivation in this work is the theoretical aim of having a minimal length, while retaining as much of the structure of quantum mechanics ({\it i.e.} the commutators between position and momentum) and of special relativity ({\it i.e.} the special relativistic dispersion relationship) as possible. 

\section{Modified energy-momentum relationship without photon dispersion}

In this section, we propose modified momentum operators which do not lead to dispersion of photons of different energies while, at the same time, having a minimal length scale. Two variants of modified momentum operators are 
\begin{equation}
    \label{tanh}
    p' = p_0 \tanh \left( \frac{p}{p_0} \right) ~~~~~~{\rm and}~~~~~~p' = \frac{2 p_0}{\pi} \arctan \left( \frac{\pi p}{2 p_0} \right)~.
\end{equation}
These two modified momentum operators have the common feature of being bounded and going over to the standard momentum at small $p$ ({\it i.e.} $p' \to p_0$ as $p \to \infty$ and $p' \to p$ for $p \ll p_0$). The physical momentum in \eqref{tanh} is $p'$. The parameter $p_0$ is the maximum momentum which characterizes the quantum gravity scale; in the following section we show that it is connected to the minimum distance. The idea of modified momentum operators having a maximum momentum, is loosely motivated by Born-Infeld electrodynamics \cite{BI} where Maxwell's equations are modified so as to have a maximum electric field strength.  

In Minkowski spacetime, one gets the dispersion relationship for a particle of mass $m$ from the energy-momentum relation $E^2 = p^2c^2 + m^2c^4$. For massless photons this energy-momentum relationship becomes $E = pc$. Using Hamilton's equation one finds $\frac{\partial E}{\partial p} = c$ so that the velocity does not depend on $E$ or $p$ and there is no dispersion. For the modified momenta, $p'$, in \eqref{tanh} we want to write down an associated modified energy, $E'$, so that in terms of $p'$ and $E'$ one has the standard relationship $E'= p'c$. Then by Hamilton's equation in terms of these modified energy and momentum one finds $\frac{\partial E'}{\partial p'} = c$ and there is no dispersion. Since the modified momentum $p'$ are bounded by $p_0$ the modified energy should also bounded by a maximum energy $E_0$. As $p \to \infty$ we want a modified energy that satisfies 
\begin{equation}
    \label{E_0}
   E ' = p' c  \to E_0 = p_0 c ~~~{\rm as}~~~ p \to \infty~.
\end{equation}
Using the two modified momentum operators from \eqref{tanh} in the energy-momentum relationship of \eqref{E_0} one finds that the modified energies are 
\begin{equation}
    \label{E' - tanh}
    E' = p_0 c~\tanh{\left(\frac{p}{p_0}\right)}  = E_0~\tanh{\left(\frac{E}{E_0}\right)}~,
\end{equation} 
and
\begin{equation}
    \label{E' - arctan}
    E' = \frac{2 p_0 c}{\pi}~\arctan{\left(\frac{\pi p}{2 p_0}\right)} =
    \frac{2 E_0}{\pi}~\arctan{\left(\frac{\pi E}{2 E_0}\right)}~.
\end{equation}
To obtain the last expressions in \eqref{E' - tanh} and \eqref{E' - arctan} we used $p=\frac{E}{c}$ and $p_0 = \frac{E_0}{c}$. 

The energy-momentum relationship between the modified energy and modified momentum as given in equations \eqref{tanh}, \eqref{E' - tanh}, and \eqref{E' - arctan} is the same as the standard energy-momentum relationship. These modified energies and momenta do not lead to dispersion since the energy-momentum relationship in \eqref{E_0} is unchanged. This lack of dispersion can be seen either from the phase velocity, $\frac{E'}{p'} = c$, or the group velocity obtained via Hamilton's equation, $\frac{\partial E'}{\partial p'} = c$. 
In contrast, several common phenomenological models of quantum gravity \cite{AC-nature,AC-PLB,AC-ijmpd,dSR,dSR1} give a modified energy-momentum relationship, like that in \eqref{AC E'} and \eqref{AC E'1}, which leads to photon dispersion as in equation \eqref{v}. 

\section{Minimal length via modified operators}

Many approaches to phenomenological quantum gravity with a minimum length scale \cite{KMM,maggiore,amati,amati2,gross,scardigli,adler-1999} use a generalized uncertainty principle (GUP). Here we review a common approach to a quantum gravity inspired minimal length given in \cite{KMM} which proposed a GUP by adding the term $\beta \Delta p^2$ to the usual Heisenberg relationship 
\begin{equation}
    \label{GUP}
    \Delta x\Delta p \ge \frac{\hbar}{2} ~\to~ \Delta x'\Delta p \ge \frac{\hbar}{2}\left (1+\beta \Delta p^2\right) ~.
\end{equation}
The phenomenological parameter $\beta$ characterizes the effect of quantum gravity and the primed quantities are modified operators. Equation \eqref{GUP} can be obtained via several different, physical approaches: using string theory \cite{vene}; using scattering arguments about high energy collision leading to black holes \cite{scardigli}; using arguments about how gravity should alter the standard Heisenberg argument for obtaining the HUP \cite{adler-1999}. In the previous approaches to a GUP one usually alters only the position operator \cite{KMM} {\it or} the momentum operator \cite{vagenas2}. Here we take the approach of altering {\it both} position and momentum operators while keeping the quantum mechanical commutator and special relativistic dispersion relationship unchanged. 

\begin{figure}[H]
    \centering
    \includegraphics[scale=0.8]{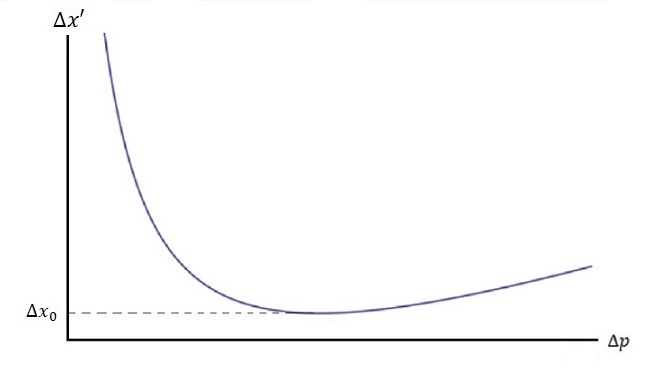}
    \caption{Generalized Uncertainty Principle from equation \eqref{GUP} where $\Delta x'$ has a local minimum of $\Delta x_0 = \hbar \sqrt{\beta}$ at $\Delta p = \frac{1}{\sqrt{\beta}}$}
\end{figure}

For the usual Heisenberg uncertainty relationship $\Delta x \propto \frac{1}{\Delta p}$, so as $\Delta p \to \infty$ one has $\Delta x \to 0$ {\it i.e.} there is no absolute, minimum length. However, from \eqref{GUP} one sees that $\Delta x' = \frac{\hbar}{2} \left( \frac{1}{\Delta p} + \beta \Delta p \right)$. Because of this $\Delta x'$ now has a local minimum, which occurs at $\Delta p = 1/\sqrt{\beta}$, and gives a minimum length of $\Delta x_0 = \hbar\sqrt{\beta}$ as shown in Fig. 1.

The addition of the $\beta \Delta p^2$ term in \eqref{GUP} can be motivated in the following way: as one probes shorter distances via higher energy/momentum collisions at some point the center of mass energy becomes great enough to create a micro black hole, whose Schwarzschild radius is set by the mass-energy in the collision {\it i.e.} $\Delta x \sim R_{sch} \sim E$. As one goes to even higher energy/momentum in the collision the resulting Schwarzschild radius will grow linearly, preventing one from probing smaller distances.   

Having a modified uncertainty relationship like \eqref{GUP} implies that there is modification of the commutator between the position and momentum operators. The usual commutator  $[{\hat x}, {\hat p}] = i \hbar$ leads to the following connection between the commutator and the HUP  
\begin{equation}
    \label{commxp}
\Delta x \Delta p \ge \frac{1}{2} \left| \langle [{\hat x}, {\hat p}] \rangle \right| \to  \Delta x \Delta p \ge \frac{\hbar}{2} ~.
\end{equation}
A GUP like that in \eqref{GUP} implies a modified commutator of the form
\begin{equation}
    \label{comm-KMM}
    [{\hat x}', {\hat p}] = i \hbar (1 + \beta p^2)~,
\end{equation}
where the primes indicate that the operators are modified. In \cite{KMM}, the position operator was modified while the momentum operator was not, and this is the reason that the $\Delta p$  and ${\hat p}$ are not primed in \eqref{GUP} and \eqref{comm-KMM}.

Modifying the commutator as in \eqref{comm-KMM} comes with some challenges as reviewed in \cite{mlake,mlake1,mlake2,mlake3}. These challenges are: (i) violation of the equivalence principle; \footnote{The violation of the equivalence principle comes from taking the classical limit of \eqref{comm-KMM} to be given by the deformed Poisson brackets, $\{ {\hat x}', {\hat p} \} = (1 + \beta p^2)$. This violation of the equivalence principle can be avoided by taking a different, more physical classical limit as explained in \cite{casadio}} (ii) the ``soccer" ball problem \cite{AC-soccer} ({\it i.e.} the difficulty in constructing consistent multi-particle states); (iii) velocity dependence of position and momentum uncertainties. In the present work we avoid these problems as well as the constraints coming from GRB observations. We do this by modifying the position and momentum operators so as to give a minimal length, while at the same time preserving the usual commutator and usual energy-momentum relationship for these modified operators.  

The usual inference drawn from the above arguments is that it is the modification of the commutator which determines the existence or not of a minimal distance. However, it was shown in reference \cite{BLS} that this is not the case; there are a host of different ways to alter the position and/or momentum operators which lead to the same commutator \eqref{comm-KMM} but may or may not give a minimal length. If instead of \eqref{GUP} one modifies the momentum operator then one would find that $\Delta x' \Delta p' \ge \frac{\hbar}{2}\left (1 + \beta {\Delta p}^2\right)$. In this case $\Delta x'$ would be proportional to  $1 / \Delta p' + \beta \Delta p^2/\Delta p'$, and whether or not this has a minimum would depend on the detailed behavior of  $\Delta p'$. The conclusion of \cite{BLS} was that it was the specific modification of the position and momentum operators which determined the existence of a minimal length rather than the way in which the commutator was modified. 

Here we show that we can use the modified momentum operators of \eqref{tanh} to obtain a minimum length scale but without modifying the fundamental commutator between the new, modified position and momentum  operator. The position operator must be modified if the commutator, in terms of the new operators, is to remain unchanged, since the momentum operators in \eqref{tanh} are changed relative to the standard momentum. In other words, we want 
\begin{equation}
\label{mod-xp}
[{\hat x}', {\hat p}'] = i \hbar ~~\to~~\Delta x' \Delta p' \ge \frac{\hbar}{2}.
\end{equation}

 We call \eqref{mod-xp} the Modified Heisenberg Uncertainty principle (MHUP) since we modify the operators but keep the commutation relationship the same, unlike the GUP where the commutator is modified. For the usual HUP, written in terms of $\Delta x$ and $\Delta p$, there is no lower bound on $\Delta x$. This is because  $\Delta x \propto \frac{1}{\Delta p}$ so $\Delta x \to 0$  as $\Delta p \to \infty$.  However, for the modified momentum operators in \eqref{tanh}, $\Delta p'$ is bounded above by $p_0$ and so $\Delta x'$ will be bounded below by $\Delta x' > \frac{\hbar}{2 p_0}$. To see the upper bound on uncertainty in modified momentum, we note that the uncertainty of the modified momentum is $\Delta p' = \sqrt{\braket{{p'}^2} - \braket{p'}^2~}$. From $\braket{p'}^2 \ge 0$ one has $\Delta p' \leq \sqrt{\braket{{p'}^2}}$. Since the magnitude of the momentum operator $p'$ is bounded above by $p_0$, we find $\braket{{p'}^2} < p_0 ^2$ and thus $\Delta p' < p_0$. In \eqref{mod-xp} $\Delta p'$ provides a smooth cut off to the momentum due to the form of the modified momentum operators in \eqref{tanh}. One could obtain the same effect on the minimum $\Delta x'$ by imposing, by hand, a hard cut-off that when $p$ reaches $p_0$ it is capped at this value. Such a hard cut-off would not preserve the special relativistic dispersion relationship and would explicitly violate Lorentz symmetry. In the concluding section we will show that the present approach of having smooth modified momentum operators, like in \eqref{tanh}, not only preserves the form of the dispersion relationship, but also maintains the form of the Lorentz generators for boosts. 

Fig. 2 shows the behavior of the MHUP as given by \eqref{mod-xp} with $\Delta p' < p_0$ and thus $\Delta x' > \frac{\hbar}{2 p_0}$. In comparing Fig. 1 with Fig. 2 one sees that although both have a non-zero minimum for $\Delta x'$ the way in which the minimum is obtained is different. Modifying the position-momentum commutator as in \eqref{comm-KMM} leads to a $\Delta x' \Delta p'$ graph which reaches a minimum $\Delta x'$ and then increases with $\Delta p'$ as shown in Fig. 1. In the present work where the position and momentum operators are modified, but the commutator in terms of the new operators is unchanged,  we find that this leads to a $\Delta x' \Delta p'$ graph which asymptotically approaches the non-zero, minimum value of $\Delta x'$ with increasing $\Delta p'$ as shown in Fig. 2. This is one of the differences between the present approach to a minimal distance and prior GUP approaches. There are  other works \cite{ghosh} \cite{nozari} which have looked at the possibility of having a minimal length while maintaining the standard commutator for modified position and momentum. 

In order to preserve the standard commutator of \eqref{mod-xp}, we need to modify the standard position operator which in momentum space is ${\hat x} = i \hbar \partial _p$. We assume that our modified position operator has the form
\begin{equation}
    \label{x' in general}
    \hat{x}'  = i\hbar f(p) \partial_p ~,
\end{equation}
and we will choose the function $f(p)$ so as to make \eqref{mod-xp} true for either modified momentum in \eqref{tanh}.

\begin{figure}[H]
    \centering
    \includegraphics[scale=0.8]{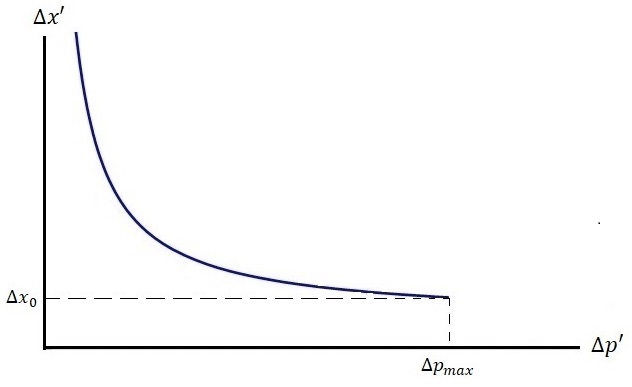}
    \caption{The relationship between $\Delta x'$ and $\Delta p'$ from the Modified Heisenberg Uncertainty Principle where $\Delta p_{max} = p_0$ and $\Delta x' > \frac{\hbar}{2 p_0}=\Delta x_0.$}
\end{figure}

Inserting the modified position \eqref{x' in general} and modified momenta \eqref{tanh} into $[{\hat x}', {\hat p'}]$ we find
\begin{equation}
    \label{CR - GHUP-tanh}
    \left[ \hat{x}' , p_0\tanh{\left(\frac{p}{p_0}\right)} \right] = i\hbar f(p) {\rm sech}^2\left(\frac{p}{p_0}\right) ~,
\end{equation}
and
\begin{equation}
    \label{CR - GHUP-arctan}
    \left[ \hat{x}' , \frac{2p_0}{\pi} 
    \arctan{\left(\frac{\pi p}{2 p_0}\right)} \right] = i\hbar f(p) \left({1+\left(\frac{\pi p}{2p_0}\right)^2}\right)^{-1}.
\end{equation}
Requiring that the right hand sides of \eqref{CR - GHUP-tanh} and \eqref{CR - GHUP-arctan} equal $i \hbar$ leads to the following modified position operators 
\begin{equation}
\label{x' -GHUP-tanh}
   \hat{x}' = i\hbar\cosh^2{\left(\frac{p}{p_0}\right)}\partial_p  ~~~{\rm for}~~~ \hat{p}' = p_0\tanh{\left(\frac{p}{p_0}\right)}
\end{equation}
and
\begin{equation}
    \label{x' - GHUP - arctan}
    \hat{x}' = i\hbar\left[{1+\left(\frac{\pi p}{2 p_0}\right)^2}\right]\partial_p  ~~~{\rm for}~~~\hat{p}' = \frac{2 p_0}{\pi}\arctan{\left(\frac{\pi p}{2 p_0}\right)}.
\end{equation}
Note that the position operator in \eqref{x' - GHUP - arctan} is that same as the modified position operator used in \cite{KMM} if one sets $\beta = \frac{\pi ^2}{4p_0 ^2}$. The modified position operator in \eqref{x' -GHUP-tanh} is reminiscent of that proposed in \cite{aiken-2019} but with $\cosh ^2 (p/p_0) \to {\rm sech} ^2 (p/p_0)$.

We have found modified position operators, modified momentum operators (given in equations \eqref{x' -GHUP-tanh} and \eqref{x' - GHUP - arctan}), and modified energies (given in \eqref{E' - tanh} and \eqref{E' - arctan}) which preserve the usual position-momentum commutator and the usual energy-momentum relationship in terms of these modified operators. Nevertheless, the modified operators lead to a minimum distance as illustrated in Fig. 2, while the energy-momentum relationship does not lead to dispersion.  Additionally, since both the commutator and energy-momentum relationship of these modified operators are the same as for the standard operators, the problems \cite{mlake,mlake1,mlake2,mlake3,AC-soccer} connected with a GUP are avoided.

In both \eqref{x' -GHUP-tanh} and \eqref{x' - GHUP - arctan} the position operator has been altered. It is in terms of these modified position operators, $\hat{x}'$, that a minimum length is obtained rather than the usual position operator, $\hat{x} = i \hbar \partial_p$. This can be seen in Figs. 1 and 2 which plot $\Delta x'$ versus $\Delta p '$ rather than $\Delta x$ versus $\Delta p '$. The question arises if this modified definition position and the related $\Delta x'$ are physically meaningful. We approach this question from the angle of ``Do the modified operators in \eqref{x' -GHUP-tanh} or \eqref{x' - GHUP - arctan} have a proper Hilbert space representation to allow them to be proper position and momentum operators?". The somewhat lengthy demonstration that modified position operators like those in \eqref{x' -GHUP-tanh} and \eqref{x' - GHUP - arctan} allow for the construction of a Hilbert space and satisfy the requirements to be physically meaningful position operators was carried out in reference \cite{KMM}. As already noted the position operator in \eqref{x' - GHUP - arctan} is exactly the same as the modified position operator used in \cite{KMM} but with $\frac{\pi^2}{4 p_0 ^2} \to \beta$. Finally in regard to the physical reasonableness of the modified position operators in  \eqref{x' -GHUP-tanh} and \eqref{x' - GHUP - arctan}, we note that in the limit $p \ll p_0$ both of the modified ${\hat x}'$s go over to the standard position operators. In the concluding section we will give estimates on lower bounds on $p_0$ coming from observations.      

\section{Modified Generalized Uncertainty Principle}

In the previous section we found a modified position, momentum and energy which gave a minimum length but without dispersion of photons of different energies, thus evading the constraints on a minimum length coming from the Fermi Gamma-Ray space telescope observations \cite{abdo}. As already mentioned the way in which the minimum distance was obtained via the GUP is different than how it was obtained via the MHUP. This is seen graphically by comparing Fig. 1 (GUP) with Fig. 2 (MHUP). For the GUP $\Delta x'$ increases linearly after reaching its minimum, whereas for the MHUP $\Delta x'$ asymptotically approaches its minimal value $\hbar \sqrt{\beta} $ as $p \to \infty$ and $p' \to p_0$. 

It is possible to retain the linear increase of $\Delta x'$ as $\Delta p'$ increases, at least for some range of $\Delta p'$. This is accomplished by picking modified position and momentum so instead of the commutator remaining unchanged, as in \eqref{mod-xp}, it takes the form
\begin{equation}
    \label{MCR - mGUP}
    \left[\hat{x}', \hat{p}' \right] = i\hbar \left(1 + \beta {p'}^2\right) ~,
\end{equation}
which is similar to the modified commutator in \eqref{comm-KMM} except here the momentum on the right hand side of \eqref{MCR - mGUP} is the modified momentum rather than the usual momentum.

The modified GUP (MGUP) \footnote{Modified GUP means modifying the both position and momentum operators, and the commutator as in \eqref{MCR - mGUP}.} associated with \eqref{MCR - mGUP} is $\Delta x' = \frac{\hbar}{2} \left( \frac{1}{\Delta p'} + \beta \Delta p' \right)$.
 Since the relationship between $\Delta x'$ and $\Delta p'$ coming from \eqref{MCR - mGUP} is exactly the same as between $\Delta x'$ and $\Delta p$ coming from \eqref{GUP} one might expect that the behavior will be identical to Fig. 1. However since \eqref{MCR - mGUP} has two parameters ({\it i.e.} $p_0$ and $\beta$) relative to \eqref{GUP} there will be two different regimes. First, when $\frac{1}{\sqrt{\beta}} \ge p_0$ the local minimum (which in Fig. 1 is at $\Delta p = \frac{1}{\sqrt{\beta}}$) will occur after the modified momentum uncertainties have reached their maximum value of $p_0$. This case is shown in Fig. 3.

\begin{figure}[H]
    \centering
    \includegraphics[scale=0.8]{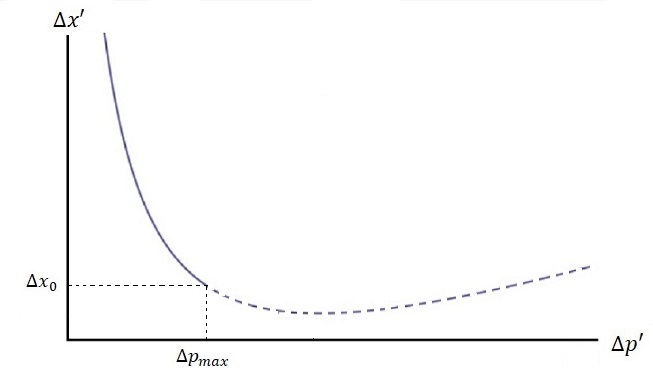}
    \caption{The modified GUP for the case when $\frac{1}{\sqrt{\beta}} \ge p_0 =\Delta p _{max}$. 
    }
\end{figure}

Second, when $\frac{1}{\sqrt{\beta}} < p_0$ the local minimum in $\Delta x'$, coming from the $\beta$ term in \eqref{MCR - mGUP}, will occur for values of $\Delta p'$ less than $p_0$. Thus $\Delta x'$ will have a local minimum before $\Delta p '$ reaches $p_0$ as shown in Fig. 4

The explicit position operators from the two modified momentum operators are generalizations of \eqref{x' -GHUP-tanh} and \eqref{x' - GHUP - arctan} and have added terms due to the $\beta$ term in \eqref{MCR - mGUP}. These modified position operators are
\begin{equation}
    \label{x' - mGUP - tanh}
\hat{x}' = i\hbar \left [\cosh^2\left(\frac{p}{p_0}\right) + \beta p_0 ^2 \sinh^2\left(\frac{p}{p_0}\right) \right]\partial_p ~~~~~{\rm for} ~~~~~ p' = p_0\tanh{\left(\frac{p}{p_0}\right)} ~,
\end{equation}
and
\begin{equation}
    \label{x' - mGUP - arctan}
\hat{x}' = i\hbar \left[{1+\left(\frac{\pi p}{2 p_0}\right)^2}\right]\left[1+ \frac{4 \beta p_0^2}{\pi^2} \arctan^2{\left(\frac{\pi p}{2 p_0}\right)}\right]\partial_p ~~~~~{\rm for}~~~~~    p' = \frac{2 p_0}{\pi}\arctan{\left(\frac{\pi p}{2 p_0}\right)}
\end{equation}

\begin{figure}[H]
    \centering
    \includegraphics[scale=0.8]{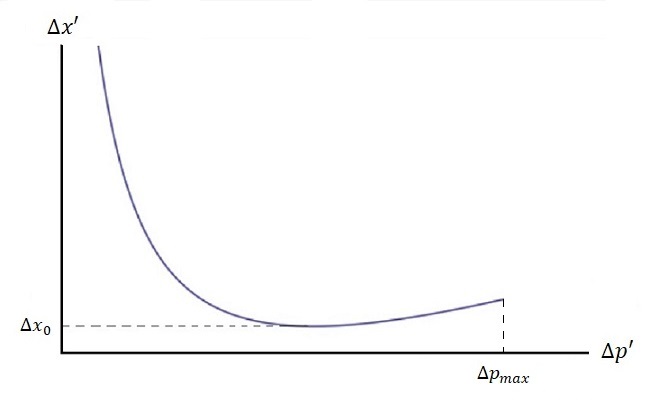}
    \caption{The modified GUP for $\frac{1}{\sqrt{\beta}} < p_0 = \Delta p_{max}$. 
    }
\end{figure}

The motivation to add the extra $\beta$ terms comes from the heuristic arguments that $\Delta x'$ should behave linearly as the energy/momentum becomes large. There are now two phenomenological parameters -- $p_0$ and $\beta$ -- which are associated with the quantum gravity scale. Recent work on modifying the uncertainty principle  \cite{mlake} also proposes two different scales. However, in \cite{mlake} the two scales were associated with a low energy/momentum scale coming from the Cosmological constant/de Sitter scale, and a high energy/momentum scale coming from the Planck/quantum gravity scale. In the above $p_0$ and $\beta$ are both associated with a phenomenological high energy/momentum scale, and they play essentially similar roles. Thus MGUP is not as economical as MHUP, since for MGUP one has two parameters, $p_0$ and $\beta$, which play essentially the same role {\it i.e.} to provide a high energy/momentum cut-off. The MGUP does lead to a linear increase of $\Delta x'$ with increasing $\Delta p'$ when $\frac{1}{\sqrt{\beta}} < p_0$ as shown in Fig. 4.

Adding the extra $\beta$ term, as in the MGUP of equation \eqref{MCR - mGUP}, opens the door to the problems associated with modifying the position-momentum commutator \cite{mlake}, namely violation of the equivalence principle, the ``soccer ball" problem, and velocity dependence of position and momentum uncertainties. This may be an argument to prefer the MHUP over the MGUP since the former does not run into these issues. 

\section{Summary and Conclusions}
 
Phenomenological, bottom up approaches to quantum gravity, like double special relativity \cite{dSR,dSR1} generically have the idea of an absolute minimal distance scale. Top down approaches to quantum gravity, such as string theory or loop quantum gravity, also incorporate the idea of a minimal distance scale. 

Generally, this absolute minimal length scale is thought to be around the Planck scale and thus hard to test. However, in \cite{AC-nature} the proposal was made that one could test for this minimal length scale using short GRBs, since the minimal distance scale would lead to an energy dependent dispersion of gamma rays  which could be detected through differences in arrival times of different energy photons. Observations of one such GRB (GRB090510) by the Fermi-Gamma ray observatory \cite{abdo} have placed constraints on this minimal distance scale to be sub-Planckian. This constraint assumed that the relationship between energy and momentum was of the general form \eqref{AC E'} and \eqref{AC E'1} leading to an energy dependence of the photon's velocity as given in \eqref{v} and a time delay \eqref{delta t}. This constraint also assumed that the lowest order correction to the photon velocity was linear in energy $E$. 

In this work we proposed modified momenta, \eqref{tanh}, and associated modified energies, \eqref{E' - tanh} and \eqref{E' - arctan} which had a maximum cut off (inspired by Born-Infeld electrodynamics \cite{BI} with its maximum electric field). However, the relationship between the modified momenta and energies, given in \eqref{E' - tanh} and \eqref{E' - arctan}, are different from the relationships given in \eqref{AC E'} and \eqref{AC E'1}. The energy-momentum relationship in \eqref{E' - tanh} and \eqref{E' - arctan} do not lead to an energy-dependent photon velocity and thus no dispersion. This avoids the constraints coming from the GRB observations. In contrast the energy-momentum relationship in \eqref{AC E'} and \eqref{AC E'1} do lead to an energy dependent photon velocity as given in \eqref{v}. 

We modified the position, momentum and energy operators in such a way that the standard position-momentum commutator and the standard energy-momentum relationship {\it had the same forms as in canonical quantum mechanics and special relativity but in terms of the modified operators}. One can ask what happens to Lorentz symmetry in the present case -- in particular what happens to boosts. The generators of boosts are $x^0 p - x p^0 = (ct) p - x (E/c)$. Since the energy and momentum are modified in the same way -- see \eqref{tanh}, \eqref{E' - tanh} and \eqref{E' - arctan} -- this implies that we should modify the time in the same way as position. For the  the $\tanh$ modification this is 
\begin{equation}
\label{time-tanh}
    \hat{x}'^0 = c\hat{t}' = i\hbar  \cosh^2\left(\frac{p}{p_0}\right)\frac{\partial}{\partial p} = i\hbar c \cosh^2\left(\frac{E}{E_0}\right)\frac{\partial}{\partial E}  ~,
\end{equation}
and for the $\arctan$ modification this is
\begin{equation}
\label{time-arctan}
    \hat{x}'^0 = c\hat{t}' =  i\hbar \left[{1+\left(\frac{\pi p}{2 p_0}\right)^2}\right]\frac{\partial}{\partial p}= i\hbar c \left[{1+\left(\frac{\pi E}{2 E_0}\right)^2}\right] \frac{\partial}{\partial E} ~.
\end{equation}
In the above we have used $E=pc$ and $E_0 = p_0 c$. With the modified position, momentum and energy one can show that $(ct') p' - x' (E'/c) = (ct) p - x (E/c)$ {\it i.e.} the form of the generators for boosts is the same as in special relativity. Using the tanh modification of energy-momentum this can be seen as follows: $(c t') p' = i\hbar c \cosh^2\left(\frac{p}{p_0}\right)\frac{\partial}{\partial p} \left[ p_0 \tanh \left( \frac{p}{p_0} \right) \right] = i \hbar c$ and for the standard operators one has $(c t) p = i\hbar c \frac{\partial}{\partial p} \left[ p \right] = i \hbar c$. Thus $(c t') p' = (c t) p$. A similar calculation applies to $x' (E'/c)$ showing that it is equivalent to $x (E/c)$. Finally for the $\arctan$ modification of energy-momentum the same type of calculation shows $(ct') p' - x' (E'/c) = (ct) p - x (E/c)$. 

The parameter $p_0$ characterizes the deviation of our modified operators from the standard operators, and it represents the energy/momentum cut-off scale in our model. One can ask what kind of constraints can be placed on $p_0$ from observations. One of the common means to place bounds on phenomenological parameters which give deviations from the standard energy-momentum and time-position operators comes from photon dispersion observation. However, these observations do not constrain $p_0$ in our model since by construction our model was intended to evade these bounds. There are other types of observations which can be used to give bounds on models which modify/violate Lorentz symmetry. These observations include pair production ($\gamma + \gamma _b \to e^- + e^+$ where $\gamma_b$ is a background photon), photon splitting ($\gamma \to \gamma + \gamma + \gamma$), suppression of air showers, and photon decay  ($\gamma  \to e^- + e^+$). There are three recent articles \cite{lamb} \cite{symmetry} \cite{rubtsov}  which discuss these constraints and give observational limits on the modification/violation of Lorentz symmetry. These constraints depend crucially on the order of the deviation in the relationship between energy and momentum. One can write a generic modified energy-momentum relationship as ${E'}^2 = p^2 + K p ^{n+2}$ which can be manipulated to yield $E' \approx p (1 + \frac{1}{2} K p^n)$. Here $K$ is some constant which characterizes ones particular model. The two common cases investigated for this modified energy-momentum relationship are $n=1$ and $n=2$. For the $n=1$ case the most stringent observational limit on the energy scale is $10^{22}$ GeV coming from photon decay. For $n=2$ the maximum limit on the energy scale is $10^{15}$ GeV coming from photon splitting. For our proposed modified operators the resulting modified energy-momentum relationships, coming  form \eqref{E' - tanh} and \eqref{E' - arctan}, are to lowest non-trivial order $E' \approx pc \left( 1 - \frac{1}{3} \frac{p^2}{p_0^2} \right)$ (to this order both $\tanh$ and $\arctan$ have the same expansion). Thus our modification corresponds to the $n=2$ case, which implies an energy bound of $10^{15} ~ {\rm GeV} < E_0$. This in turn gives the bound on $p_0$ of $10^{15} ~ {\rm GeV/c} ~ < p_0$. 

We also obtained a minimal distance via a modified Heisenberg Uncertainty Principle ({\it i.e.} modification of the operators but not the commutator) in contrast to obtaining a minimal distance via the standard Generalized Uncertainty Principle ({\it i.e.} modification of the commutator {\it and} operators).
In conjunction with the modified momenta from \eqref{tanh}, we also introduced appropriately modified position operators, that led to a minimal distance. This was accomplished without modifying the fundamental position -momentum commutation relationship. This avoids many of the problems and issues associated with modifying the position-momentum commutator \cite{mlake,AC-soccer}. The approach to a minimal distance presented here also leads to a different relationship between the uncertainty in position and the uncertainty in momentum, as compared to the usual GUP approach, as can be seen by comparing Figs. 1 and Figs. 2.

There is recent work \cite{kuntz} which uses the Schwinger-Keldysh formalism to investigate the existence or not of a minimum geometrical length versus a minimum length scale. This work comes to the conclusion that while there is no minimum geometrical length, there is a minimum length scale (identified as the Planck scale) beyond which scattering experiments become useless. Reference \cite{kuntz} also suggests, as is the case in the present work, that quantum gravity effects can not be probed by looking for dispersion of photons as they propagate through spacetime. This distinction between a minimum length scale versus a minimum geometrical length was presaged in earlier works \cite{amati,amati2,gross,scardigli,adler-1999,vene} which make similar arguments.

\end{document}